\documentclass{epl}

\title{Vanishing Fe 3d orbital moments in single-crystalline magnetite}
\author{E. Goering\inst{1} \and S.Gold\inst{1} \and M.Lafkioti\inst{1} \and G.Sch\"utz\inst{1} }
\institute{
  \inst{1} Max-Planck-Institut f\"ur Metallforschung, Heisenbergstrasse 3, 70569 Stuttgart, Germany\\
  }

\begin{document}

\maketitle

\begin{abstract}
We show detailed magnetic absorption spectroscopy results of an {\it in situ} cleaved high quality single crystal of magnetite. In addition the experimental setup was carefully optimized to reduce drift, self absorption, and offset phenomena as far as possible. In strong contradiction to recently published data, our observed orbital moments are nearly vanishing and the spin moments are quite close to the integer values proposed by theory.  This very important issue supports the half metallic full spin polarized picture of magnetite. 
\end{abstract}

\section{Introduction}

Magnetite \chem{Fe_3O_4} has been fascinating mankind for thousands of years \cite{Ref1}. \chem{Fe_3O_4} shows a phase transition, the so called Verwey transition at {\it T$_V$} $\approx$ 123K, accompanied with a jump in the electrical conductivity, which has been extensively investigated in the last century (For a review see Ref. \cite{Ref2} or \cite{Ref3} and references therein). Today magnetite has attracted enormous interest, because of the proposed high-spin-polarization and related possible applications for future spin-electronic-devices \cite{Ref4}. The nature of the conducting electrons and the influence of local electronic correlations are one of the key issues to understand \chem{Fe_3O_4}\cite{Ref5}. 
\chem{Fe_3O_4} crystallizes at room temperature in the antiferromagnetic cubic inverse Spinel structure ($Fd\overline{3}m$), formally written as \chem{Fe(A)Fe(B)_2O_4} \cite{Ref6}. The A-type ions are tetrahedrally coordinated and nominally in a Fe$^{3+}$ ($\approx$ -5$\mu_B$) configuration. The B-site ions are located on octahedral sites and mixed valent with equally distributed Fe$^{3+}$ ($\approx$ +5$\mu_B$) and Fe$^{2+}$ ($\approx$ +4$\mu_B$) ions. The magnetic moments shown in brackets are pure spin moments related to a fully occupied local majority band (opposite for A and B sites). The magnetic moments of the A and B sites are aligned antiparallel to each other with a resulting magnetization per formula unit 5$\mu_{B (B)}$+4$\mu_{B (B)}$-5$\mu_{B (A)}$= 4$\mu_B$, consistent to the experimental result of 4.07$\mu_B$ \cite{Ref7}. An observation of an integer spin moment is therefore a clear indication for a B-site minority electron conduction mechanism, and its accompanied full spin polarization at the Fermi level. The phase transition at {\it T$_V$} = 123K been has explained by Verwey in terms of a charge localization-delocalization of the conducting B-site electrons \cite{Ref8,Ref9,Ref10}. This discussion is recently revived experimentally and theoretically by refined structural data results \cite{Ref11,Ref12,Ref13}, which found only a slightly corrugated charge order between 2.4-2.6e, accompanied by orbital ordering \cite{Ref14,Ref13}.\\

In a recent Letter \cite{Ref15} the spin and orbital Fe 3d magnetic moments of magnetite have been evaluated experimentally by X-ray magnetic circular dichroism (XMCD) and calculated within the LDA+U scheme using a rotationally invariant LDA+U functional. Non integer spin moments and large unquenched B-site orbital moments of 0.33$\mu_B$ have been found, which have been attributed to a strong onsite Coulomb interaction and corresponding 3d correlation effects. Other reported band structure calculations and LDA+U calculations, using the original LDA+U functional, found only  small orbital moments, and the Fe spin moments were just slightly reduced from the integer value due to hybridization with O \cite{Ref16,Ref17,Ref18,Ref5}.\\

In this contribution, we will show carefully performed XMCD experiments, which are consistent to the majority of theoretical predictions  \cite{Ref16,Ref17,Ref18,Ref5}, confirming the integer moment description and the full spin polarized model of \chem{Fe_3O_4} with a very small integral orbital moment. The extracted sum rule \cite{Ref19,Ref20} related  spin and orbital moments are therefore in contradiction to Ref. \cite{Ref15}. To clearly point out the quality and validity of our results, we will discuss in detail the experimental procedure and possible experimental error sources, like surface effects, signal drift, self absorption, and offset phenomena. The chosen experimental setup has been optimized to minimize these error sources as far as possible.

\begin{figure}[ttt]
\centering
\onefigure[width=10cm]{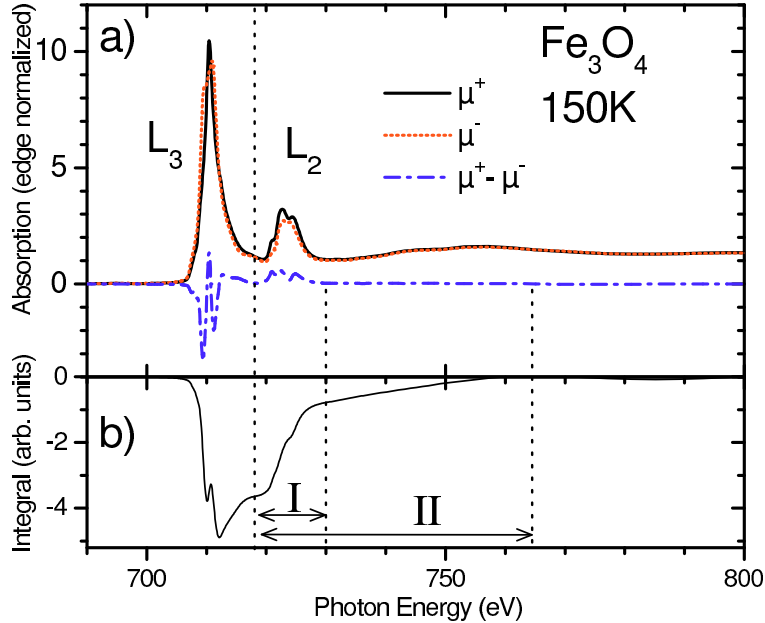}
\caption{a) X-ray absorption and XMCD spectrum of a fractured single crystal of magnetite at 150K and normal incidence geometry. b) Integrated XMCD spectrum used for sum rule analysis. Different sum rule integration ranges are indicated as I and II.}
\label{Fig01}
\end{figure}

\section{Experimental}

All XMCD spectra have been measured in total electron yield mode (TEY) at the BESSY II bending magnet beamline PM3 in an applied magnetic field of 10kOe, providing full sample saturation at all temperatures. All sum rule related values have been corrected for the finite circular polarization of 0.93$\pm$0.02. The absorption has been normalized to the incoming photon beam intensity by measuring synchronous the photocurrent of a gold grid. To prevent even very small non XMCD related current asymmetries, normal incidence conditions have been used and the sample was biased at -100V to minimize magnetic field induced electron backscattering phenomena \cite{Ref21}. The residual so called raw data "XMCD-like-offset", which is not related to the intrinsic magnetic properties of the sample, was smaller than 1/1000 of the total absorption. But even this very small residual XMCD-like-offset, monitored at energies below 690eV, has been numerically treated and subtracted. This procedure and the physics behind the "XMCD-like-offset" have been described in detail elsewhere \cite{Ref21}.\\
The sample magnetization has been flipped at every data point to compensate even for smallest synchrotron related drift phenomena. In addition, the results have been verified for opposite light helicity. The monochromator energy resolution was set to $\approx$ 6000. High quality synthetic single crystalline sample have been prepared by V.A.M. Brabers in an arc-image furnace using the floating zone technique \cite{Ref22}. Single crystals have been annealed and cooled under equilibrium conditions to obtain highly stochiometric single crystals \cite{Ref23}. The vacancy concentration of the used \chem{Fe_{3-x}O_4} single crystal is smaller than x $<$ 10$^{-6}$, proven by measuring magnetic after effect spectra \cite{Ref2,Ref24}. As derived from SQUID measurements, the Verwey temperature of the single crystal is {\it T$_V$}  = 124K. The single crystal has been {\it in situ} cleaved at room temperature (surface normal along [110] direction), at an ambient pressure of smaller than 10$^{-9}$ mBar.\\
We would like to emphasize that our cleaved sample has shown - during the same synchrotron measurement time - a sharp shift of the O K- edge onset monitoring the gap variation at the Verwey transition \cite{Ref25}. This demonstrates that the shown absorption results of the cleaved single crystal used here, exhibit bulk properties of magnetite. This is also consistent to recently published high energy photoemission results \cite{Ref26}. \

\begin{figure}[ttt]
\centering
\onefigure[angle=270, width=10cm]{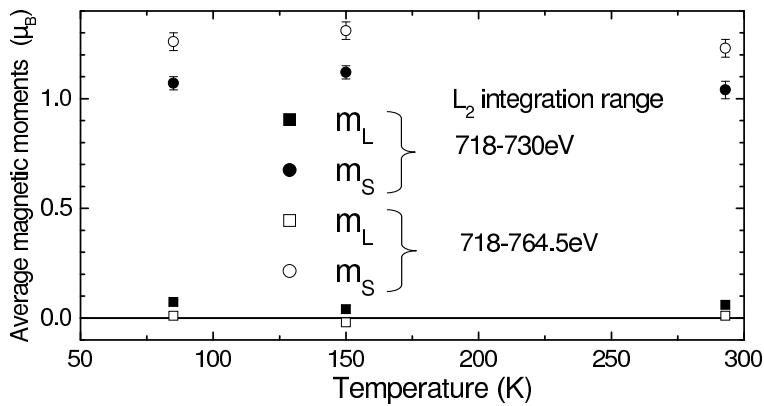}
\caption{Spin and orbital magnetic moments, determined for a short and extended integration range at different temperatures (see text).}
\label{Fig02}
\end{figure}

A typical XAS and XMCD spectrum is shown in Fig.1a. The general shape of the XAS lines is comparable to the results presented in Ref. \cite{Ref15}. The integrated XMCD signal (Fig. 1b) exhibits a non vanishing slope between 730-765eV, which is related to a very small and reproducible XMCD signal above the L$_2$ edge region.\\

Sum rules \cite{Ref20,Ref19} have been applied for two different L$_2$  edge integration ranges (range I: 718-730eV and range II: 718-764.5eV) to show the influence of this small high energy XMCD signal. The background intensity has been subtracted in a standard way, with L$_3$  (L$_2$) edge centre position at 710.7eV (723.3eV) \cite{Ref27}, and the same number of Fe 3d holes N$_h$/Fe=13.5/3 has been used as in Ref. \cite{Ref15}. Results are shown in Fig.2. For the orbital moments the estimated error bar is about the symbol height. Only minor temperature dependencies are observable, comparable to SQUID related magnetization curves. An important point is the absence of prominent difference between above and below the Verwey transition. 
The average orbital moments are -0.001$\mu_B$ (0.06$\mu_B$) for the extended (short) integration range. The obtained average sum rule related effective spin moment/FU is ${( m_s + 7<T_z>)=3.90 \pm 0.09\mu_B}$ (long range). 

\begin{figure}[hhh]
\centering
\onefigure[angle=270,width=10cm]{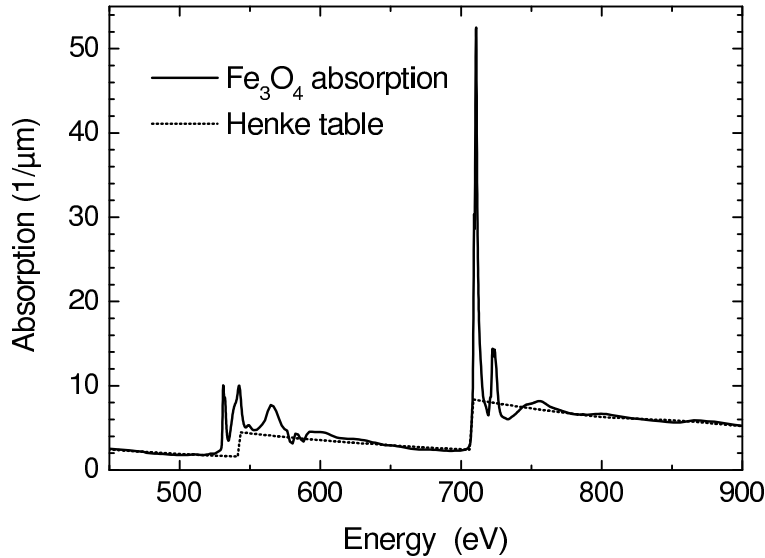}
\caption{\chem{Fe_3O_4} nonmagnetic normal incidence TEY signal adapted and compared to the Henke table tabulated absorption \cite{Ref32}.}
\label{Fig03}
\end{figure}

\section{Discussion}
 
The necessity using such a long integration range can be explained by the presence of magnetic extended fine structure (MEXAFS) oscillations at the L$_{2,3}$ edges, which are always superimposed to the resonant XMCD structure \cite{Lemke,Ahlers}. Therefore, some part of the L$_3$ edge MEXAFS is superimposed to the L$_2$ edge signal. This modifies sum rule values with short integration range. Due to the fact that the MEXAFS signal is strongly damped at higher excitation energies, the dominating parts of the L$_3$ and L$_2$ edges, close to the excitation threshold, have opposite sign and cancel each other \cite{Ahlers} for extended integration ranges. A similar relative strong MEXAFS related intensity has been identified for \chem{CrO_2}, another ferromagnetic oxide \cite{GoeringPRL}.\\
 
The observed nearly vanishing orbital moment is consistent to the pure LDA calculation in ref. \cite{Ref15} and to other theoretical approaches \cite{Ref16,Ref17,Ref18,Ref5}. 
The effective spin moment/FU of $(m_s +7<T_z>)=3.90 \pm 0.09\mu_B$ is much larger than the uncorrected LDA+U value of 3.2$\mu_B$ from Ref. \cite{Ref15}, but in perfect agreement with the calculation of Penicaud {\it et al}. \cite{Ref17}, and quite close to other reference values \cite{Ref16,Ref18,Ref5}. The result presented here exhibits a nearly integer spin value per formula unit and supports the full spin polarized picture of magnetite as introduced above. Due to the predominantly cubic symmetry, even below $T_v$, and because of the consistency of our results, we do not believe that $T_z$ plays an important role in our measurements \cite{Stoehr}.       

At this point we will quantitatively discuss saturation effects, to make obvious the validity of our results. In the past S. Gota {\it et al}. have been investigated \chem{Fe_3O_4}TEY mode saturation effects for a thin film sample as a function of the angle of incidence and sample thickness \cite{Ref29}, where an unusual large value of $\lambda_e$ = 5nm has been found, which is about 2-7 times larger compared to other published effective electron escape length values \cite{Ref30,Ref31,Ref32,Ref33}. \\ 

\begin{figure}[hhh]
\centering
\onefigure[width=7cm]{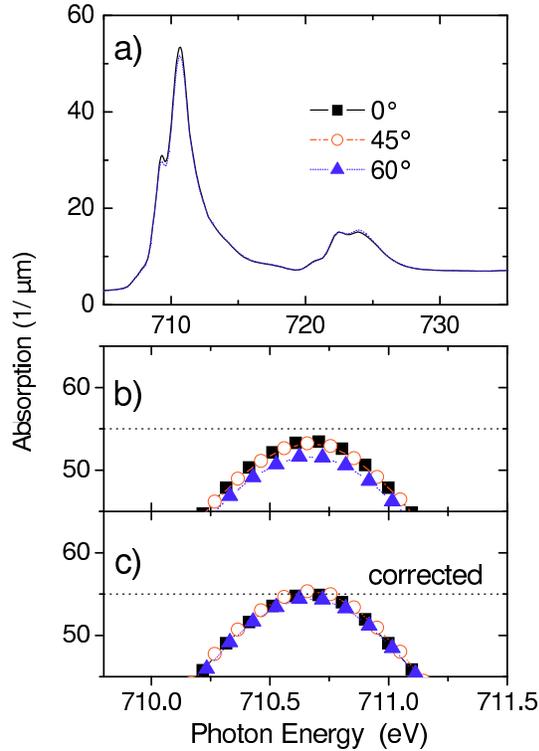}
\caption{a) Angular dependent nonmagnetic Henke table adapted TEY spectra of a polished single crystal. Maximal self absorption effects are shown in an expanded view: Uncorrected (b) and corrected (c).}
\label{Fig04}
\end{figure}

Therefore, we have performed angular dependent XAS measurements on a polished [111] \chem{Fe_3O_4}surface, from a fracture of the single crystal discussed above. Figure 3 shows a long energy range normal incidence spectrum, which is carefully adapted to the Henke table atomic absorption curves for \chem{Fe_3O_4} \cite{Ref34,Ref35}. The determined resonant absorption length scales are quite similar but also slightly reduced, compared to the absorption published by S. Gota {\it et al}. \cite{Ref29}. This reduction, not significant for the conclusions given here, is related to the extended adaptation range, which we have used to reduce errors related to the oscillating region between 730-780eV \cite{Ref30}.\\
Figure 4a shows the Fe L$_{2,3}$ edge region of the Henke table adapted nonmagnetic absorption lines for normal (0 $^\circ$\hspace{-2pt}C) and 60 $^\circ$\hspace{-2pt}C angle of incidence. Only very small differences are observable at high absorption values near 710eV. This region is magnified in Figure 4b for 0 $^\circ$\hspace{-2pt}C, 45 $^\circ$\hspace{-2pt}C, and 60 $^\circ$\hspace{-2pt}C of incidence. We corrected these absorption coefficients for self absorption (Fig. 4c), choosing an effective electron escape length of $\lambda_e$=0.8$\pm$0.3nm \cite{Ref32}, in order to achieve best agreement between the corrected angular dependent curves. This single crystal related effective electron escape length is much smaller compared to the $\lambda_e$ =5nm thin film result \cite{Ref29}. A very similar value of 0.7nm has been previously determined for another oxide system: \chem{CrO_2} \cite{Ref30}.  

The total peak height variation induced by saturation effects is about 3\% (5.4\%) for the normal incidence (60$^\circ$\hspace{-2pt}C) spectrum (difference between Fig 4b and 4c). Due to the oscillating behaviour of the L$_3$ edge XMCD the influence to the orbital moment is even smaller. In case of the fractured single crystal result at normal incidence, the orbital moment saturation correction is smaller than 0.01$\mu_B$. In order to check the maximum influence of the saturation effects, we have additionally calculated the orbital moment variation for the large escape length $\lambda_e$ = 5nm \cite{Ref29}, where the orbital moment shift is only 0.05$\mu_B$. The determined orbital moment per B site is therefore less than 3/2*0.01$\mu_B$=0.015$\mu_B$ (even for $\lambda_e$ =5nm smaller than 0.08$\mu_B$). The different escape length between this contribution and the results from Gota {\it et al}. can be in principle related to different surface roughness parameters, where roughness gives rise to an effective variation of the angle of incidence. On the other hand open micro (or nano) trenches could also be responsible to an enhanced effective electron escape length. Up to now this has not been investigated on a quantitative basis. Due to the lack of surface related information, we will not discuss the difference between both escape length estimations. It is above the scope of this contribution to discuss all other possible explanations for the observed differences in the effective electron escape length, for example electrical conductivity variations.\\ 
Nevertheless, we have clearly shown that for normal incidence geometry saturation effects do not significantly modify our XMCD results, even for the large electron escape length values from Gota {\it et al}.\
Our results demonstrate the necessity to obtain high quality wide range XMCD data for \chem{Fe_3O_4} to extract correct orbital moment values. However, even for the short integration range the orbital moment is far below the values presented in Ref. \cite{Ref15}. The origin of this difference is most likely not related to the physics of \chem{Fe_3O_4}, and will be discussed in more detail elsewhere \cite{Ref28}.

\section{Conclusion}

In conclusion, our carefully performed experimental XMCD results for magnetite are in very good agreement to the majority of theoretical investigations. Self absorption and other experimental effects have been neglected as a possible source of significant XMCD related moment variation, especially for the orbital moment. Taking into account an extended XMCD integration range gives a magnetic spin moment, quite close to the expected integer value of 4$\mu_B$, while the orbital moment vanishes. This result clearly supports the full spin polarized picture of magnetite.

\acknowledgments
We thank M. F\"ahnle, F. Walz, and H. Kronm\"uller for fruitful discussions, and the Max-Planck-Society for financial support. Many thanks to V.A.M.Brabers providing the magnetite sample and T. Kachel from BESSY II for support during the beamtime.

\end{document}